\documentclass[twocolumn,preprintnumbers,nofootinbib]{revtex4}
\usepackage{graphicx}
\usepackage{amssymb}
\usepackage{color}
\usepackage{enumerate}
\def\be{\begin{equation}}
\def\ee{\end{equation}}
\def\bea{\begin{eqnarray}}
\def\eea{\end{eqnarray}}

\renewcommand{\texttt}{{}}

\begin{document}

\title{Ho\v{r}ava-Lifshitz theory as a Fermionic Aether in Ashtekar gravity}

\author{Stephon Alexander}
\email{salexand@haverford.edu}
\affiliation{Department of Physics and Astronomy, Dartmouth College,
Hanover, NH, USA}
\author{Jo\~ao Magueijo}
\email{j.magueijo@imperial.ac.uk}
\affiliation{Theoretical Physics, Blackett Laboratory, Imperial College, London, SW7 2BZ, UK}

\author{Antonino Marcian\`o}
\email{amarcian@haverford.edu}
\affiliation{Department of Physics, The Koshland Integrated Natural Science Center, Haverford College, Haverford, PA 19041 USA} 
\affiliation{Department of Physics, Princeton University, New Jersey 08544, USA}

\date{\small\today}

\begin{abstract} \noindent
We show how Ho\v{r}ava-Lifshitz (HL) theory appears naturally in the Ashtekar 
formulation of relativity if one postulates the existence of a fermionic
field playing the role of aether.  The spatial currents associated
with this field must be switched off for the equivalence to work. Therefore
the field supplies the preferred frame associated with
breaking refoliation (time diffeomorphism) invariance, but obviously
the symmetry is only spontaneously broken if the field is dynamic. 
When Dirac fermions couple to the gravitational field via the Ashtekar 
variables, the low energy limit of HL gravity, recast in the language
of Ashtekar variables, naturally emerges (provided the spatial 
fermion current identically vanishes). HL gravity can therefore be
interpreted as a time-like current, or a Fermi aether, that fills space-time,
with the Immirzi parameter, a chiral fermionic coupling, and the fermionic 
charge density fixing the value of the parameter $\lambda$
determining HL theory. 
This reinterpretation sheds light on some features
of HL theory, namely its good convergence properties. 
\end{abstract}

\maketitle

\section{Introduction} 

\noindent While there are stringent experimental constraints on breaking local Lorentz invariance in particle physics, it is well known that diffeomorphism invariance plays a more prominent structural role in general relativity and quantum gravity since it is possible that near the Planck scale, Lorentz symmetry is not fundamental.   One of our best tests of Lorentz invariance on large distance scales is the CMB, which breaks Lorentz invariance by choosing a preferred time-like frame for the Universe during the epoch of last-scattering.  Given this fact, one may be tempted to construct gravitational theories that have a preferred frame from the outset while preserving diffeomorphism invariance. But what more is there to gain from working with gravitational theories that violate Lorentz-invariance?

  Recently, some authors have constructed theories of gravity that have preferred-frame effects ({\it i.e.} an Einstein Aether), but preserve spatial-diffeomorphisms.  One of the attractive features of a class of these models, namely Ho\v{r}ava-Lifschitz Gravity (HL) \cite{Horava:2009uw}, is that, due to their anisotropic scaling, implementation of standard field theory methods renders the UV behavior of gravity perturbatively finite.  Therefore in this scheme, Lorentz invariance can emerge in the IR, but its violation at shorter scales can cure the UV infinities that usually plague perturbative general relativity.

Despite the promise that HL gravity provides, breaking of refoliation invariance has led to certain technical issues, most notably the presence of an extra scalar graviton mode \cite{HII}.  The theory could certainly be improved with the import of extra ingredients coming from other walks of gravitational theory. It is interesting that the discreteness of space-time in Loop Quantum Gravity (LQG) also provides a natural UV regulator \cite{T} and one is led to wonder if the finiteness in HL gravity is connected to the non-perturbative discreteness found in LQG.  A way to begin analyzing this possible connection is to see if HL gravity can be reexpressed in terms of the Asthekar canonical variables which naturally lead to the the holonomy representation of LQG. 

In this paper we show that HL gravity can indeed be reexpressed in terms of Ashtekar's variables and a new physical interpretation of the HL theory emerges, which paves a way of understanding a manifestly $4D$ formulation of HL without the need for an extra scalar degree of freedom.   What we will discover is that when Dirac fermions couple to the gravitational field via the Ashtekar variables, HL gravity emerges when the spatial fermion current identically vanishes. Vanishing of fermionic currents in equivalent physical systems has been considered {\it e.g.} in \cite{Giacosa:2008rw} and \cite{Alexander:2009uu}, and we refer to these works for a detailed analysis. For us it is interesting to note that the 
frame in which this happens supplies the ``preferred'' foliation of the theory. In Ho\v{r}ava-gravity the finiteness of the graviton arises due to the presence of the Cotton-Tensor which was assumed.  In this work we discover a physical reason for this in the Ashtekar variables:  when the condition for the York-time \cite{Y2} is imposed, the extrinsic curvature gets related to the Cotton tensor and the York time is identified with the zeroth component of the fermion current, i.e. the charge density.  In this phase, HL gravity has the interpretation of a time-like current (Fermi-Aether) that fills space-time.   We also show an equivalence of the scalar, vector and Gau\ss\ constraints between HL gravity and the Ashtekar constraints when the spatial fermion current vanishes.

\section{HL theory in Ashtekar variables}\label{hlash}

\noindent 
One cannot overemphasize the importance of spinors in understanding
gravity and its quantization. Starting from Weyl, it was understood that
the simplest way to couple spinors to gravity involved the so-called
spin-connection, in the ``Cartan-Palatini'' formulation of general relativity.
Later Kibble realized that general relativity could be seen as the gauge
theory of the Poincar\'e group, with the tetrad gauging translations
and the spin-connection gauging Lorentz transformations and rotations. 
Torsion naturally sneaks
into the theory whenever spinors are present, although the relation is
purely algebraic, so that torsion can 
be reinterpreted as a 4-fermion interaction in the 
standard torsion-free theory (for an excellent review see~\cite{hammond} 
and reference therein).

To a large extent the Ashtekar formalism is a reformulation of the
Palatini-Cartan-Kibble earlier work, rendering it more amenable to
quantization via techniques imported from lattice gauge theory. 
The Ashtekar theory can be obtained by adding a surface term to the usual 
Palatini action. Depending on how this is done in the spinorial sector, one may 
end up with the same classical dynamics or with an extension of the original
theory when spinors are present, as we shall see
in the next Section. In either case the
quantum theory is always distinct from what one would get by attempting 
to quantize the original theory. Quantum effects and classical dynamics 
driven by spinors always introduce novelties.


One may wonder how the HL theory 
looks using Ashtekar's ``new'' variables. This is most easily
accomplished following the treatment in~\cite{T}, where the Ashtekar formalism
is derived from the standard ADM framework by an extension of the
phase space followed by a canonical transformation (dependent
on the Immirzi parameter $\gamma$). The first operation
produces a canonical pair made up of the densitized inverse triad
$E^a_i$ and the extrinsic curvature 1-form $K^i_a$~\footnote{From now on, we will label space indices with latin letters $a, b$, with $a,b=1,2,3$, and internal SU$(2)$ indices with latin letters $i,j$, with $i,j=1,2,3$.} . With
$E^i_a$ the inverse of $E^a_i$, the extrinsic curvature $K_{ab}$ can be 
obtained from the ``extended'' $K^i_a$ according to:
\be
K_{ab}=\sqrt{q}K^i_{(a}E^i_{b)}
\ee
subject to constraint:
\be
G_{ab}=K^i_{[a}E^i_{b]}=0
\ee
(which produces a form of the Gau\ss\ constraint when contracted with 
$\epsilon^{cab}$).
A canonical transformation dependent on Immirzi parameter $\gamma$
is then applied to $K^i_a$ leading to the Ashtekar connection:
\be
A^i_a=\gamma K^i_a +\Gamma ^i_a\; ,
\ee
where, {\it in the absence of spinors}, $\Gamma^i_a=\tilde{\Gamma}^i_a$ is the torsion-free
Cartan connection associated with $E^a_i$. The Gau\ss\ constraint
implies
$D_aE^a_i=\partial_a E^a_i +\epsilon_{ijk}\Gamma^j_aE^a_k=0$, which leads
to an expression in terms of the new covariant derivative:
\be
{\cal G}_i={\cal D}_aE^a_i=\partial _a E^a_i +\epsilon _{ijk} A^j_a E^a_k=0\; .
\ee
This is the usual form for the Gau\ss\ constraint in terms of 
Ashtekar variables. The Gau\ss\ constraint is the only new constraint to be
added in this approach to the usual two present in the ADM formalism.

Having performed this exercise, the ADM Hamiltonian becomes the
sum of 3 constraints: the Gau\ss\, the diffeormorphism and the Hamiltonian 
constraint. Specifically the
Hamiltonian constraint becomes:
\be
{\cal H}_{\rm Ash}=
\frac{1}{2\kappa\sqrt{q}}{E}^{a}_{i}E^{b}_{j}\left(\epsilon^{ij}_{\,\, k}F_{ab}^{k}-2(\gamma^{2}+1)K_{[a}^{i}K_{b]}^{ j}\right) \; ,
\ee
(where we are using units such that $\kappa=8\pi G$). 

We now note that the HL action can be written as the standard 
Einstein-Hilbert action plus an additional term in $1-\lambda$:
\be \label{achl}
S_{\rm HL}= S_{\rm EH}+\frac{1-\lambda}{2\kappa}\int d^3 x dt \sqrt{q} NK^2\; .
\ee
This results in a correction to the ADM Hamiltonian:
\be
{\cal H}_{\rm HL}={\cal H}_{\rm ADM} +\frac{\sqrt{q}}{2\kappa}(\lambda -1) K^2\; .
\ee
Therefore all we need to do in order to translate the model into the Ashtekar 
formalism is to rewrite the extra term in terms of the canonically
transformed variables. 
It is easy to prove that:
\be
K=q^{ab}K_{ab}=\frac{1}{\sqrt{q}}E^a_iK^i_a\; ,
\ee
so that the Hamiltonian constraint becomes
\bea
{\cal H}_{\rm HL} &=&
\frac{1}{2\kappa\sqrt{q}}{E}^{a}_{i}E^{b}_{j}\Big(\epsilon^{ij}_{\,\, k}F_{ab}^{k}-2(\gamma^{2}+1)K_{[a}^{i}K_{b]}^{ j}\nonumber \\
&&+ (1-\lambda) K_{a}^{i}K_{b}^{ j}\Big) \; .
\eea
We see that the diffeomorphism invariant theory contains both the trace and the traceless part in well apportioned amounts. The new term is a pure trace, deforming the original proportions. Notice finally that when we select the values $\lambda=1+2(\gamma^2+1)$, the theory becomes:
\bea \label{zione}
{\cal H}_{\rm HL} &=&
\frac{1}{2\kappa\sqrt{q}}{E}^{a}_{i}E^{b}_{j}\Big(\epsilon^{ij}_{\,\, k}F_{ab}^{k}+ 2 (1-\lambda)K_{[a}^{i}K_{b]}^{ j}\nonumber \\
&&+ (1-\lambda) K_{(a}^{i}K_{b)}^{ j}\Big) \; .
\eea

Our task now is to obtain this theory from a fermionic aether. 
In so doing it will be useful to recall that in the
above Hamiltonian $K^i_a$ is to be understood as
\be
K^i_a=\frac{A^i_a-\Gamma^i_a}{\gamma}\; .
\ee
Thus, if $\Gamma^i_a$ acquires torsion (solved explicitly in terms 
of the fermionic field), it is not unreasonable to expect that 
a new term, of the form of the new term in $(1-\lambda$), is generated.

\section{Einstein-Hilbert action and coupling to massless fermions} \label{echn}

\noindent A direct way to see how HL gravity is related to the Ashtekar variables is to consider a $4D$ gravitational Holst action in the first-order formalism which can naturally be reduced to the Ashtekar variables \cite{M,AV,PR,FMT,BD}:
\bea\label{new action}
\!\!\!\!\!\!\!\!&& S_{EHC}\left(e,A,\psi,\overline{\psi}\right)= \\
\!\!\!\!\!\!\!\!&&=\frac{1}{2 \kappa}\int_\mathcal{M} \left(\frac{\epsilon_{IJKL}}{2}\,e^{I}\wedge e^{J}\wedge F^{KL}-\frac{1}{\gamma}\,e_{I}\wedge e_{J}\wedge F^{IJ}\right) +\nonumber
\\
\!\!\!\!\!\!\!\!&&+\frac{i}{2}\!\int_{\mathcal{M}}\!\!\! \!\star e_I  \wedge \! \left[\overline{\psi}\gamma^I \! \left(1-\frac{i}{\alpha}\gamma_5\right) \!\mathcal{D}\psi-\overline{\mathcal{D}\psi}\left(1-\frac{i}{\alpha}\gamma_5 \! \right)\!\gamma^I\psi\right], \nonumber
\eea
in which anti-symmetrized pairs $AB$, with $A,B=0,1,2,3$, are internal indices of the adjoint representation of $\mathfrak{so}(3,1)$ and the symbol $\mathcal{D}$ denotes covariant derivative with respect to the $SO(3,1)$ connection $A^{IJ}$, the field strength of which is $R^{IJ}$. Notice that this action differs from the one considered in \cite{PR} by an axial coupling in the fermionic term.  It was shown in~\cite{M} that this action is equivalent to the Einstein-Cartan action at the effective level.  We can immediately identify the the Ashtekar-Barbero connection as a spatial projection of the spin-connection: 
\begin{equation}
A^{'j}_{b} \equiv -\gamma A_{b}^{\
  j0}-\frac{1}{2}\epsilon^{j}_{\  kl} A_{b}^{\  kl}= 
\gamma K_b^j+\Gamma_b^j 
\end{equation}
(where both sets of indices run from 1 to 3, as previously stated).
The remaining components of the space-time connection $A$ are recast into:
\begin{equation}
{\,^{-}\!\!A}^{'j}_{b}\equiv A_{b}^{\
  j0}-\frac{1}{2\gamma}\epsilon^{j}_{\  kl}A_{b}^{\  kl}. 
\end{equation}
Finally the components $A_t^{IJ}$ are non-dynamical, as are the lapse function $N$ and shift vector $N^a$ appearing in the metric. Variation with respect to the non-dynamical connection components gives partially second class constraints. These constraints can be solved, giving the results 
\begin{eqnarray}
\label{gammabk2}
\gamma {\,^{-}\!\!A}_{b}^{' k}&=&-A_{b}^{' k}+ 2{\Gamma}^{k}_{b}\,.
\end{eqnarray}
Following \cite{BD}, we rewrite the connection $\Gamma^{k}_{b}$ as 
\begin{eqnarray}
\Gamma^{k}_{b}&=& \widetilde{\Gamma}^{k}_{b}+\frac{\gamma\kappa}{4(1+\gamma^{2})}
\left(\theta  \ \epsilon_{ij}^{\ \  k}e_{b}^{i}\mathcal{J}^{j}-\beta e_{b}^{k} \mathcal{J}^{0}\right)\,, 
\end{eqnarray} 
{\it i.e.} the sum
of the metric compatible spin connection $\widetilde{\Gamma}^k_b$ and 
a torsion contribution
\begin{eqnarray}
\label{cbk1}
C^{j}_{a} 
&\equiv & \frac{\gamma
  \kappa}{4(1+\gamma^{2})}\left(\theta \ \epsilon^{j}_{\
    kl}e_{a}^{k}\mathcal{J}^{l}- \beta e_{a}^{j} \mathcal{J}^{0}\right)\,,
\end{eqnarray}
with coefficients
\be
\beta=\gamma+\frac{1}{\alpha}\, \quad {\rm and} \quad \theta=1-\frac{\gamma}{\alpha}\; ,
\ee
where the currents are defined as
\be
\mathcal{J}^0=\phi^\dag\, \phi- \chi^\dag \, \chi\,, \qquad \mathcal{J}^i =\phi^{\dagger}\sigma^{i}\phi+\chi^{\dagger}\sigma^{i}\chi\,,
\ee
in terms of the spin components $\psi\!=\!(\phi, \chi)^T$. Furthermore, $A_t^{k0}$ is determined by another second class constraint, requiring $\epsilon_{ijk}A_t^{jk}$ to remain free as Lagrange multiplier of the Gau\ss\, constraint. 

With the definitions above the Gau\ss\ constraint becomes:
\begin{equation}
\label{uno}
\mathcal{G}_i  = \gamma [K_{b},E^{b}]_{i} -\frac{\gamma \beta}{2(1+\gamma^{2})}\sqrt{q}\mathcal{J}_{i}\, .
\end{equation}
The diffeomorphism constraint reads:
\begin{eqnarray}\label{due}
&&\mathcal{C}_a = \frac{1}{\gamma}E^{b}_{j}F_{ab}^{j}- \frac{i}{\gamma} \sqrt{q}\left(\theta_{L}(\phi^{\dagger}
{D}_{a}{\phi}-
\overline{{D}_{a}{\chi}}\,\chi)- c.c.\right)+ \nonumber\\
&&-
\frac{\gamma^{2}+1}{\gamma^2}K^{j}_{a}G_{j} \,,
\end{eqnarray}
while the Hamiltonian constraint is:
\begin{eqnarray}\label{tre}
&&\mathcal{C} = 
\frac{1}{2 \kappa \sqrt{q}}{E}^{a}_{i}E^{b}_{j}\left(\epsilon^{ij}_{\
  \ k}F_{ab}^{k}-2(\gamma^{2}+1)K_{[a}^{i}K_{b]}^{ j}\right)+ \nonumber\\
&&+\frac{ \beta}{2 \kappa \gamma\sqrt{q}}{E}_{i}^{a}{\Delta}_{a}(\sqrt{q}\mathcal{J}^{i})+(1+\gamma^{2})
\kappa\widetilde{D}_{a}
\left(\frac{{E}_{i}^{a}G^{i}}{\sqrt{q}}\right)  \nonumber \\ 
&& +\frac{i}{\gamma\kappa}
E^{a}_{i}\Big(\theta_{L}(\phi^{\dagger}\sigma^{i}{\Delta}_{a}\phi+\overline{{\Delta}_{a}\chi}\sigma^{i}\chi))+\nonumber\\
&& -\theta_{R}(\chi^{\dagger}\sigma^{i}{\Delta}_{a}\chi+\overline{{\Delta}_{a}\phi}\sigma^{i}\phi)\Big) + \nonumber\\
&& +\frac{1}{4 \kappa \gamma^2} \left(3-\frac{\gamma}{\alpha}+
2\gamma^{2}\right) \epsilon_{lkr} K_{a}^{l}E^{a}_{k}\mathcal{J}^{r}\,,
\end{eqnarray}
where $D$ is the covariant derivative with respect to $\Gamma^k_b$, $\tilde{D}$ is the covariant derivative with respect to compatible connection $\tilde{\Gamma}^k_b$, and we have introduced $\theta_{L/R}\equiv \frac{1}{2}(1\pm i/\alpha)$. The derivative $\Delta$ stands for the covariant derivative related to the ``corrected connection'' $\mathcal{A}^i_a$ (see Ref.~\cite{BD} for a detailed description), whose expression in terms of the connection $A^i_a=\tilde{A}^i_a+\bar{A}^i_a$ accounting for the torsion-full components $\bar{A}^i_a$ is given by 
\be
\mathcal{A}^i_a \equiv {A}^{i}_a + \frac{\gamma \kappa}{4 \alpha} e^i_a \mathcal{J}^0\,,
\ee
where $\tilde{A}^i_a={A'}^i_a$. In the notation of \cite{BD} the Ashtekar-Barbero connection splits into a torsion part and a torsion-free part. Specifically, with $\tilde{\Gamma}^i_a$ the compatible torsion-free spin-connection and $\tilde{K}^i_a$ the compatible torsion-free extrinsic curvature, we have:
\be
A^{i}_a=\tilde{\Gamma}^i_a+ \gamma \tilde{K}^i_a + \frac{\kappa \gamma}{4}\, \epsilon^i_{\,\,kl}\, e^k_a\, \mathcal{J}^l - \frac{\kappa \gamma}{4 \alpha} e^i_a\,  \mathcal{J}^0\,.
\ee
The three constraints (\ref{uno})--(\ref{tre}) provide a set of first class constraints.

\section{Non-minimal ECH action in metric-compatible variables} 

\noindent We focus on the term in $\mathcal{C}$, as its generalization introduces us to the Hamiltonian formulation of the Ho\v{r}ava-Lifshitz dynamics. The Gau\ss\, and the vector constraints of the Einstein-Cartan-Holst action will indeed close weakly on the constraints' surfaces, the same constraints' algebra where the Ho\v{r}ava-Lifshitz theory of gravity (\ref{achl}) closes, provided some extra conditions are satisfied. In contrast, the Ho\v{r}ava-Lifshitz term in (\ref{achl}), in the scalar constraint, $\mathcal{H}_{\rm HL}$,  endows the De Witt metric with a conformal dimensionless coupling $\lambda$.  For $\lambda< 1/3$  gravity becomes repulsive, and it is interesting to notice that this condition corresponds to a region in the plane spanned by the Immirzi parameter $\gamma$ and non-minimal Fermion coupling constant $\alpha$.  

We start from the scalar constraint for the Einstein-Cartan-Holst action (\ref{new action}) recast in terms of the ``metric compatible'' Ashtekar variables:
\bea \label{ali}
&&\mathcal{H}^{\rm ECH}_{\rm Ash}=\frac{1}{2\kappa\sqrt{q}}{E}^{a}_{i}E^{b}_{j}\left(\epsilon^{ij}_{\,\, k}F_{ab}^{k}-2(\gamma^{2}+1)K_{[a}^{i}K_{b]}^{ j}\right) + \nonumber\\
&&+\frac{i}{2 \gamma} E^a_i (\phi^\dagger \sigma^i \partial_a \phi - \chi^\dagger \sigma^i \partial_a \chi -c.c.) +\nonumber\\
&&+ \frac{\theta}{2 \gamma} E^b_j \tilde{\Gamma}^j_b \mathcal{J}^0 + \frac{\gamma }{4 \alpha \sqrt{q}} \epsilon^{ij}_{\,\,\,\,k} E^a_i e^k_b \mathcal{J}^0 \partial_a E^b_j + \nonumber\\
&& + \frac{3  \kappa}{16} \frac{\sqrt{q}}{1+\gamma^2} \left(  \frac{1}{\alpha^2} -\frac{2}{\alpha \gamma} -1 \right) \left( \mathcal{J}_0^2 - \mathcal{J}_l \mathcal{J}^l \right) + \nonumber\\ 
&& + \frac{1}{\kappa \gamma^2} \tilde{D}_a \left( \frac{E^a_i \tilde{G}^i}{\sqrt{q}} \right) + \frac{2+\gamma^2}{4 \gamma^2}  \tilde{G}_i \mathcal{J}^i \,,
\eea
where the tilde ``$\tilde{\phantom{a}}$'' labels metric-compatible quantities. We will  show that it is possible to reduce $\mathcal{H}^{\rm ECH}_{\rm Ash}$  to the Ho\v{r}ava-Lifshitz gravity scalar constraint 
\bea \label{orali}
{\cal H}_{\rm HL} &=&
\frac{1}{2\kappa\sqrt{q}}{E}^{a}_{i}E^{b}_{j}\Big(\epsilon^{ij}_{\,\, k}F_{ab}^{k}-2(\gamma^{2}+1)K_{[a}^{i}K_{b]}^{ j}\nonumber \\
&&+ (1-\lambda) K_{a}^{i}K_{b}^{ j}\Big) \,,
\eea
by assuming some restrictions on the quantum states of the Fermionic matter content of (\ref{new action}). In order to show the equivalence of the two theories, we must also check that the vector constraint $\mathcal{C}_a$  and the Gau\ss\, constraint $\mathcal{G}_i$, once recast in the metric-compatible variables, reduce to the ones of the Ho\v{r}ava-Lifshitz theory provided some assumptions (that will soon be listed) are fulfilled.  

We first rewrite the Gau\ss\ and Vector constraints in terms of metric-compatible quantities.  In the presence of fermions the Gau\ss\ constraint is modified to  
\be
\mathcal{G}_i= D_b E^b_i - \frac{1}{2} \sqrt{q} \mathcal{J}_i = \gamma [K_b,E^b]_i -\frac{\gamma \beta}{ 2 (1+\gamma^2)} \sqrt{q} \mathcal{J}_i
\,.
\ee
We see that when the spatial current vanishes ($\mathcal{J}_l=0$) the Gau\ss\, constraint reduces to $\mathcal{G}_i=\gamma \epsilon^k_{\,\,j i} K^k_b E^b_j= \gamma \epsilon^k_{\,\,j i} \tilde{K}^k_b E^b_j = \tilde{\mathcal{G}}_i$. 
We can express the vector constraint in terms of metric-compatible variables as
\bea
&&\mathcal{C}_a=\frac{1}{\gamma}\, E^b_j \tilde{D}_{[a} \tilde{K}_{b]}^j + {\rm sign\, (det} e^i_a) \, \frac{\kappa}{4}\, \epsilon_{c a}^{\,\,\,\,b}\, E^c_l \tilde{D}_b (\sqrt{q} \mathcal{J}^l) + \nonumber\\
&&- \frac{i}{2 \gamma} \sqrt{q} \left(\phi^\dagger\tilde{D}_a \phi + \chi^\dagger\tilde{D}_a \chi - c.c. \right) +\nonumber\\
&&+ \frac{1}{\gamma} {\rm sign\, (det} e^i_a)\, E^d_l (\epsilon_{c d}^{\,\,\,b} \Gamma^c_{ba} - \epsilon_{c a}^{\,\,\,b} \Gamma^c_{bd}  ) \sqrt{q} \mathcal{J}^l+\nonumber\\
&&+\left( \frac{\kappa}{4} \epsilon^{jkl} \mathcal{J}_k e_{al} - \frac{\kappa}{4 \alpha} e^j_a \mathcal{J}^0 - \frac{1+\gamma^2}{\gamma} K^j_a \right) \tilde{\mathcal{G}}_j\,,
\eea
which reduces to the expression
\bea
\mathcal{C}_a&\!=\!&\frac{1}{\gamma}\, E^b_j \tilde{D}_{[a} \tilde{K}_{b]}^j -\left(\frac{\kappa}{4 \alpha} e^j_a \mathcal{J}^0 + \frac{1+\gamma^2}{\gamma} K^j_a \right) \tilde{\mathcal{G}}_j \nonumber\\
&\!=\!& \tilde{\mathcal{C}}_a -  \left(\frac{\kappa}{4 \alpha} e^j_a \mathcal{J}^0 + \frac{1+\gamma^2- \gamma^3}{\gamma} K^j_a \right) \tilde{\mathcal{G}}_j 
\eea
on states over which $\mathcal{J}^l$ and $(\phi^\dagger\tilde{D}_a \phi + \chi^\dagger\tilde{D}_a \chi - c.c.)$ vanish. Once we classically implement $\mathcal{G}_i\!=\!\tilde{\mathcal{G}}_i\!=\!0$, it follows that the vector constraint of the Einstein-Cartan-Holst theory (\ref{new action}) becomes equivalent  to the one expressed in terms of the metric compatible variables in Ho\v{r}ava-Lifshitz gravity.  For this to be true it is essential that the spatial currents remain switched off (for a discussion of this condition see {\it e.g.} Refs.~\cite{Giacosa:2008rw, Alexander:2009uu}).

\subsection{Fixed values of the Immirzi parameter and $\lambda$}

\noindent In this Section we explore the case $\theta\!=\!0$, emphasizing that the equality it encodes between the two parameters entering the Einstein-Cartan-Holst action ({\it i.e.} $\alpha\!=\!\gamma$) is not necessarily required. Indeed $\alpha$ does not need to be fixed to $\gamma$ for an equivalence between the Ho\v{r}ava-Lifshitz theory of gravity endowed with the square of the Cotton tensor, namely the term $C_{ij} C^{ij}$, and the action in (\ref{new action}) to be found. Nevertheless, we start from this instructive and simple case. Throughout this Section, we assume that\footnote{We denote with ``$\langle \, \cdot \rangle$'' the expectation value of operators on the quantum state realizing our assumptions.} $\langle(\phi^\dagger\tilde{D}_a \phi + \chi^\dagger\tilde{D}_a \chi - c.c.)\rangle$ and $\langle\mathcal{J}_i\rangle$ vanish, as a necessary condition for our claim. We then recast the theory in terms of torsion-full Ashtekar variables $\{A^i_a, E^b_j\}$, which in turn are given, in terms of the compatible variables $\{\tilde{A}^i_a, \tilde{E}^b_j\}$, by
\be
A^i_a= \tilde{A}^i_a - \frac{\gamma \kappa}{4 \alpha} e^i_a \mathcal{J}^0\,, \qquad     E^b_j=\tilde{E}^b_j \,.
\ee
This relation yields general extrinsic curvature, $K^i_a$, which has a torsion-free part, $\tilde{K}^i_a$, and a torsion-full piece $\overline{K}^i_a$
\be \label{K}
K^i_a= \tilde{K}^i_a - \frac{\kappa}{4 \alpha} e^i_a \mathcal{J}^0 \equiv \tilde{K}^i_a +  \overline{K}^i_a \,,
\ee
where the torsion-full extrinsic curvature is $\overline{K}^i_a=-\frac{\kappa}{4 \alpha} e^i_a \mathcal{J}^0$.

As we are rewriting our theory in terms of torsion-full quantities, for the sake of consistency the field-strength must be expressed in terms of the torsion-full connection $A^i_a$.  It is not difficult to check that the scalar constraint (\ref{ali}) re-writes as
\bea \label{baba}
\!\!\!\!&&\mathcal{H}^{\rm ECH}_{\rm Ash}=  \frac{1}{2 \kappa \sqrt{q}}{E}^{a}_{i}E^{b}_{j}\left(\epsilon^{ij}_{\ \
    k}(F_{ab}^{k}- 2(\gamma^{2}+1)K_{[a}^{i}K_{b]}^{ j}\right)\nonumber\ \\  
\!\!\!\!&&+\frac{i}{2 \kappa \gamma} E^a_i \,(\phi^{\dagger}\sigma^{i}\widetilde{D}_{a}\phi-
  \chi^{\dagger}\sigma^{i}\widetilde{D}_{a}\chi-c.c.) + \nonumber\ \\  
\!\!\!\!&&+\frac{ {E}_{i}^{a}}{2\kappa\,\sqrt{q}}\widetilde{D}_{a}(\sqrt{q}\mathcal{J}^{i})
  +\frac{1}{2 \kappa}{E}_{j}^{b}K_{b}^{j}\mathcal{J}^{0} + \frac{1}{2 \kappa \gamma}[K_{a},E^{a}]_{j}\mathcal{J}^{j} +
 \nonumber\ \\ 
 \!\!\!\!&&
  -\frac{3}{8 \kappa \sqrt{q}}\frac{1}{1+\gamma^{2}}q\, \mathcal{J}_{0}^2+ \frac{1+\gamma^{2}}{\kappa \gamma^2 } \, \widetilde{D}_{a}
\left(\frac{{E}_{i}^{a}\mathcal{G}^{i}}{\sqrt{q}}\right)\,.
\eea
Provided now that $\langle \mathcal{J}^i \rangle=\langle (\phi^{\dagger}\sigma^{i}\widetilde{D}_{a}\phi-
  \chi^{\dagger}\sigma^{i}\widetilde{D}_{a}\chi-c.c.)  \rangle=0$, on the constraint's surface  where $\mathcal{G}^{i}=0$ the scalar constraint (\ref{baba}) written in terms of torsion-full quantities reads 
\bea \label{hec3}
\!\!\mathcal{H}^{\rm ECH}_{\rm Ash} \!\!&=&\!\!  \frac{1}{2 \kappa \sqrt{q}}{E}^{a}_{i}E^{b}_{j}\left(\epsilon^{ij}_{\ \
    k} F_{ab}^{k}- 2(\gamma^{2}+1)K_{[a}^{i}K_{b]}^{ j}\right) + \nonumber\ \\  
\!\!&-&\!\!\frac{3}{8 \kappa \sqrt{q}}\frac{1}{1+\gamma^{2}}q\, \mathcal{J}_{0}^2\,, .
\eea
A few algebraic manipulations are now in order. Firstly note that
\bea \label{rimani}
\frac{3 \kappa}{16} \frac{\sqrt{q}}{\gamma^2} \mathcal{J}_0^2= \frac{2}{3} \,\frac{E^a_i E^b_j}{ 2 \kappa \sqrt{q} } \left( \frac{\kappa}{4\gamma} e^i_{(a} \,  \frac{\kappa}{4\gamma} e^j_{b)}  \right) \mathcal{J}_0^2 \,,
\eea
in which the symmetrization arises from the fact that
\bea 
9 e^2\!\!=\!\! (E^a_i e^i_a ) (E^b_j e^j_b)\!=\! E^a_i E^b_j\!  \left(  e^i_{(a} e^j_{b)} + e^i_{[a} e^j_{b]} \right)\!\!=\!\! E^a_i E^b_j\, e^i_{(a} e^j_{b)}, \nonumber
\eea
where symmetrization and skew-symmetrization are intended to have been normalized (recall too that $\sqrt{q}=e$). It is then straightforward to recognize that 
\bea
\frac{2}{3} \,\frac{E^a_i E^b_j}{ 2 \kappa \sqrt{q} } \left( \frac{\kappa}{4\gamma} e^i_{(a} \,  \frac{\kappa}{4\gamma} e^j_{b)}  \right) \mathcal{J}_0^2= \frac{2}{3} \, \frac{E^a_i E^b_j}{ 2 \kappa \sqrt{q} }\, \overline{K}^i_{(a} \overline{K}^j_{b)}    
\eea
and that
\bea
\!\!\!\!&& \frac{2}{3} \,\frac{E^a_i E^b_j}{ 2 \kappa \sqrt{q} } \left( \frac{\kappa}{4\gamma} e^i_{a} \,  \frac{\kappa}{4\gamma} e^j_{b}  \right) \mathcal{J}_0^2=  \nonumber\\
\!\!\!\!&& = \frac{2}{3} \, \frac{E^a_i E^b_j}{2 \kappa \sqrt{q} }\, \left( K^i_{a} K^j_{b}  + \tilde{K}^i_{a} \tilde{K}^j_{b} -2 K^i_{a} \tilde{K}^j_{b} \right) \,,
\eea
having made use of the definition of $\overline{K}^i_a$ in (\ref{K}) and again
the identitities $E^a_i e^i_a =3 e$ and $E^a_i=e \, e^a_i$. If we impose that the trace of the extrinsic curvature vanishes, $K\!=\!0$, which in terms of metric-compatible variables, is equivalent to imposing
\be \label{York time}
\widetilde{K}=-\frac{3}{4 \gamma} \, \mathcal{J}_0\,,
\ee
we obtain
\be
\frac{3 \kappa}{16} \frac{\sqrt{q}}{\gamma^2} \mathcal{J}_0^2= \frac{2}{3} \, \frac{E^a_i E^b_j}{2 \kappa \sqrt{q} }\, \left( K^i_{a} K^j_{b}  + \tilde{K}^i_{a} \tilde{K}^j_{b} \right)\,.
\ee
We emphasize that condition (\ref{York time}) (which generalizes the Lichnerowicz condition $\widetilde{K}\!=\!0$) corresponds to the second class constraint imposed to the ADM formulation of gravity $\Pi\!=\!\mathcal{Y}$  ($\Pi^{ab}$ being the conjugate momentum to $q_{ab}$) while solving the vector and scalar constraints. The York time\footnote{More precisely, we should consider the definition of the York time provided in \cite{Y2} (and recalled in \cite{IshamTime}), in which the trace of $\Pi^{ab}$ is rescaled by the inverse of $\sqrt{q}$ in order to provide a variable canonically conjugated to the Hamiltonian density $\sqrt{q}$. It is not probably surprising the fact that this automatically encodes a treatment of fermonic matter in terms of densitized fields (see {\it e.g.} Refs.~\cite{wbc, T, AMS}). } $\mathcal{Y}$ is then identified with the fermionic electric charge density:
\be
\mathcal{Y}=-\frac{3}{4 \gamma} \, \mathcal{J}_0\,.
\ee

Once these algebraic manipulations are considered, 
it immediately follows that 
\bea \label{hec4}
 \!\!\mathcal{H}^{\rm ECH}_{\rm Ash}\!\!&=&\!\! \frac{1}{2 \kappa \sqrt{q}}{E}^{a}_{i}E^{b}_{j}\Big(\epsilon^{ij}_{\ \  k} F_{ab}^{k}- 2(\gamma^{2}+1)\,K_{[a}^{i}K_{b]}^{ j}+  \nonumber\\
\!\!&-&\!\! \frac{2}{3} \frac{\gamma^2}{1+\gamma^2}\, \overline{K}_{(a}^{i} \overline{K}_{b)}^{ j} \Big) \,.
\eea
Therefore, when $\lambda=3+ 2 \gamma^{2}$ and $3(\gamma^2+1)^2\!=\!\gamma^2$, which respectively fix the values $\gamma^2=\{-3,-1/3\}$ and $\lambda=\{-3, 7/3\}$, we find that the scalar constraint for action (\ref{new action}) is equivalent
to the scalar constraint of the Ho\v{r}ava-Lifshitz gravity theory, {\it i.e.}
\bea \label{EHC-HL}
\!\!\!\!\!\!\!\!\!\!\!\!&&\mathcal{H}^{\rm ECH}_{\rm Ash} \!\!=\!\!  \frac{1}{2 \kappa \sqrt{q}}{E}^{a}_{i}E^{b}_{j}\Big(\epsilon^{ij}_{\ \
    k} F_{ab}^{k}- 2(\gamma^{2}+1)K_{[a}^{i}K_{b]}^{ j} + \nonumber\ \\  
 \!\!\!\!\!\!\!\!\!\!\!\!\!&&+ (1-\lambda) K_{a}^{i}K_{b}^{ j}   \Big) 
\; ,
\eea
with 
\be
\lambda\!=\! 3+ 2 \gamma^{2}\!=\!\{-3,{7}/{3}\}\,. 
\ee
In terms of the torsion-full variables, the Gau\ss\, and the vector constraint becomes: 
\be
\mathcal{G}_i =  D_bE^b_i -\frac{1}{2} \sqrt{q}\, \mathcal{J}_i=
\gamma [K_b,\,E^b]_i - \frac{\gamma \beta}{2(1+\gamma^2)} \sqrt{q} \mathcal{J}_i
\ee
and
\bea
\!\!\!\!\!\!&&\mathcal{C}_a=\frac{1}{\gamma}E^b_j F^j_{ab} -\frac{1+\gamma^2}{\gamma^2} K^i_a - \frac{i}{2 \gamma} \sqrt{q} \left(\phi^\dagger \widetilde{D}_a \phi + \chi^\dagger \widetilde{D}_a \chi \right) +\nonumber\\
\!\!\!\!\!\!\!\!&& -\frac{ \kappa\,\sqrt{q} }{8(1+\gamma^2)}\! \left( \theta \epsilon^j_{\ kl} e^k_a \mathcal{J}^l-\beta e^j_a \mathcal{J}_0 \right)\! \mathcal{J}_i -\frac{1}{2} K^i_a \sqrt{q} \mathcal{J}_i \,.
\eea
Again, under the assumptions $\langle \mathcal{J}^i \rangle=\langle (\phi^{\dagger}\sigma^{i}\widetilde{D}_{a}\phi-
  \chi^{\dagger}\sigma^{i}\widetilde{D}_{a}\chi-c.c.)  \rangle=0$, we recover that $\mathcal{G}_i$ and $\mathcal{C}_a$ have the same form as the Gau\ss\, and the vector constraints of the Ho\v{r}ava-Lifshitz theory of gravity, provided that torsion-free quantities are replaced everywhere by torsion-full quantities.  
  
It is remarkable that when the York-time condition is imposed, $K\!=\!0$, the Cotton tensor is naturally present in the theory. Indeed, as shown in \cite{Tiemblo:2005vi} using Ashtekar variables, under the assumption $K\!=\!0$ the constraints imply
\be \label{KkC}
\widetilde{K}^{ab}=k \, \varepsilon^{abd} \,\widetilde{D}_a \left( \widetilde{R}_d^{\ b}-\frac{1}{4} \delta^b_d \widetilde{R} \right)=k\,\widetilde{C}^{ab}\,,
\ee
with $\widetilde{R}_a^{\ b}$ the three-dimensional Ricci tensor and $\widetilde{R}$ its contraction, $\varepsilon^{abd}$ the Levi-Civita tensor $\varepsilon^{abd}=\epsilon^{ijk}\,e^a_i e^b_j e^c_k$, $k$ a constant of proportionality and $\widetilde{C}^{ab}$ the Cotton tensor in 3D in terms of metric-compatible variables. 
We recall that the action for the $z=3$ Ho\v{r}ava-Lifshitz theory of gravity  in $3+1$D  takes the form
\bea \label{WHL}
\!\!\!\!S^{\rm HL}\!=\!\!\! \int \!\!dt d^3x\! \sqrt{q} N \!\left(\! \frac{2}{\kappa^2} (\widetilde{K}_{ij} \widetilde{K}^{ij} \!-\! \lambda \widetilde{K}^2) \!-\! \frac{\kappa^2}{2 w^4} \widetilde{C}_{ij} \widetilde{C}^{ij} \!\!\right)\!, 
\eea
which after Wick rotation to imaginary time may be re-written as a sum of squares
\bea
&&S^{\rm HL}=2 i \int \!\! dt d^3x \sqrt{q} N \left( \frac{1}{\kappa} \widetilde{K}_{ij}  - \frac{\kappa}{2 w^2} \widetilde{C}_{ij} \right)
G^{ijkl} \times \nonumber\\
&& \times \left( \frac{1}{\kappa} \widetilde{K}_{kl}  - \frac{\kappa}{2 w^2} \widetilde{C}_{kl} \right)\,,
\eea
where we have introduced the de Witt metric 
\be
G^{ijkl}= \frac{1}{2} \left( q^{ik} q^{jl} + q^{il} q^{jk}  \right) - \lambda q^{ij} q^{kl} \,.
\ee
When we impose (\ref{York time}), we find that (\ref{KkC}) relates the metric compatible extrinsic-curvature and the metric-compatible Cotton tensor. Therefore, the two tensors depend on the extrinsic curvature terms that only appear in the scalar constraint (\ref{EHC-HL}) and in action (\ref{achl}). This would finally account for recovering an action similar in form to (\ref{WHL}) after having properly Wick rotated the time coordinate, but with contribution originated by the presence of fermions (and consequently of torsion). On shell, for solutions of the Hamiltonian constraints derived from (\ref{new action}) once (\ref{York time}) is imposed, a relation similar in form to (\ref{achl}) can be recovered, but now in terms of torsion-full quantities. The Wick-rotated action is then 
\bea
 S\!=\!2 i \!\!\int \!\! dt d^3x \sqrt{q} N \Big[ \frac{1}{\kappa} \left(\widetilde{K}_{ij}  + \overline{K}_{ij}   \right) 
G^{ijkl} \frac{1}{\kappa} \left(\widetilde{K}_{kl}  + \overline{K}_{kl}   \right) \!\Big]\,. \nonumber  
\eea
By introducing real parameter $\xi$ and using
$\widetilde{K}^{ab}=k\,\widetilde{C}^{ab}$ we can write
the Euclidean action as
\bea
 &&S=2 i \!\!\int \!\! dt d^3x \sqrt{q} N \Big\{\Big[ \frac{1-\xi}{\kappa} \left(\widetilde{K}_{ij}  + \overline{K}_{ij}   \right) 
G^{ijkl} \times \nonumber\\
&& \times \frac{1-\xi}{\kappa} \left(\widetilde{K}_{kl}  + \overline{K}_{kl}   \right) \Big]+ \frac{\xi^2}{\kappa^2} \widetilde{C}_{ij} \widetilde{C}^{ij} +  2 \xi \, \widetilde{K}_{ij} G^{ijkl} \overline{K}_{kl} \Big\}\, . \nonumber  
\eea
This finally becomes
\bea
 S\!\!&=&\!\!2 i \!\!\int \!\! dt d^3x \sqrt{q} N \Big\{ \frac{1}{{\kappa'}^2} K_{ij} G^{ijkl} K_{ij}  + \frac{\xi^2}{(1-\xi)^2{\kappa'}^2} \widetilde{C}_{ij} \widetilde{C}^{ij}  \nonumber\\
\!\!\!\!\!\!&-&\!\!  2 \frac{\xi^2}{(1-\xi)^2{\kappa'}^2} \, \overline{K}_{ij} G^{ijkl} \overline{K}_{kl} \Big\}\,, \nonumber  
\eea
once we recognize that $\widetilde{K}_{ij}G^{ijkl} \widetilde{C}_kl$ can be written as a total derivative \cite{Horava:2009uw}, for parameters
\bea \label{lista}
\!\!\frac{\xi^2\, k^2}{(1-\xi)^2}\!=\!\frac{{\kappa'}^4}{2 w^2}\,\,\,\,{\rm and}\,\,\,\, (\gamma^2\!, \lambda)\!=\!\left\{(-3,-3),(-\frac{1}{3},\frac{7}{3})\right\}\!,
\eea
and absorbing  $1-\xi$ in $\kappa$, by defining
\bea
\kappa'=\kappa/(1-\xi)\,. \nonumber
\eea
At the end of this procedure, we obtain the Euclidean action
\bea \label{quasi}
 S\!\!&=&\!\!2 i \!\!\int \!\! dt d^3x \sqrt{q} N \Big\{ \frac{1}{{\kappa'}^2} K_{ij} G^{ijkl} K_{ij}  + \frac{{\kappa'}^2}{2 w^2} \widetilde{C}_{ij} \widetilde{C}^{ij}  \nonumber\\
\!\!\!\!\!\!&-&\!\! \frac{(1-3\lambda)}{\gamma^2}  \frac{{\kappa'}^4}{ w^2} \, \frac{3(1-\xi)^2}{16} \mathcal{J}_0^2 \Big\}\,. 
\eea
Action (\ref{quasi}) contains a $\mathcal{J}_0^2$ interaction-term additional to the action for the Ho\v{r}ava-Lifshitz gravity in \cite{Horava:2009uw}. Without fixing $\lambda$ the last term in (\ref{quasi}) can be made to vanish for the degenerate value $\lambda=1/3$ (instead of (\ref{lista})). By properly dealing with $\alpha$ more general solutions can be found, and the conditions relating $\gamma$ and $\lambda$ and the vanishing of the $\mathcal{J}_0^2$ terms can be met simultaneously. We will revisit this issue in the next sub-section. 
We close this section with a remark on the two possible values of $\gamma^2$, and hence $\lambda$, which we have found were needed for our equivalence. It is well known~\cite{Giulini:1994dx} that the physical meaning of $\lambda$ can be inferred from the analysis of the acceleration of the three-volume $V\equiv \int d^3x \sqrt{q}$, which is encoded in the formula
\be
\frac{d^2}{dt^2} V= -\frac{2}{3\lambda -1} \, \int d^3x \sqrt{q} \widetilde{R}\,.
\ee
Therefore, an attractive gravitational force is recovered for $\gamma^2=-1/3$ and $\lambda=7/3$ in this framework.

\subsection{The of $\alpha$-parameter solutions in HL gravity }

\noindent In this sub-section we show how it is possible to extend our  results, dropping the constraint $\alpha=\gamma$, leading to a one-parameter family of solutions in $\lambda$ and $\gamma$ depending on the non-minimal coupling $\alpha$ entering the Einstein-Cartan-Holst action (\ref{new action}). We will show that it is possible to impose the vanishing of the extra interaction term in (\ref{quasi}) even for $\alpha\neq\gamma$.
The scalar constraint will still be given by
\bea \label{alibaba}
\!\!\!\!&&\mathcal{H}^{\rm ECH}_{\rm Ash}=  \frac{1}{2 \kappa \sqrt{q}}{E}^{a}_{i}E^{b}_{j}\left(\epsilon^{ij}_{\ \
    k}(F_{ab}^{k}- 2(\gamma^{2}+1)K_{[a}^{i}K_{b]}^{ j}\right)\nonumber\ \\  
\!\!\!\!&&+\frac{i}{2 \kappa \gamma} E^a_i \,(\phi^{\dagger}\sigma^{i}\widetilde{D}_{a}\phi-
  \chi^{\dagger}\sigma^{i}\widetilde{D}_{a}\chi-c.c.) + \nonumber\ \\  
\!\!\!\!&&+\frac{ {E}_{i}^{a}}{2\kappa\,\sqrt{q}}\widetilde{D}_{a}(\sqrt{q}\mathcal{J}^{i})
  +\frac{1}{2 \kappa}{E}_{j}^{b}K_{b}^{j}\mathcal{J}^{0} + \frac{1}{2 \kappa \gamma}[K_{a},E^{a}]_{j}\mathcal{J}^{j} +
 \nonumber\ \\ 
 \!\!\!\!&&
  -\frac{3}{8 \kappa \sqrt{q}}\frac{1}{1+\gamma^{2}}q\, \mathcal{J}_{0}^2+ \frac{1+\gamma^{2}}{\kappa \gamma^2 } \, \widetilde{D}_{a}
\left(\frac{{E}_{i}^{a}\mathcal{G}^{i}}{\sqrt{q}}\right)\,,
\eea
but now the definition of the torsion-full part of the extrinsic curvature $\overline{K}^i_a=-\frac{\kappa}{4\alpha} e^i_a \mathcal{J}^0$ allows us to re-express the scalar constraint as
\bea \label{hec5}
 \!\!\mathcal{H}^{\rm ECH}_{\rm Ash}\!\!&=&\!\! \frac{1}{2 \kappa \sqrt{q}}{E}^{a}_{i}E^{b}_{j}\Big(\epsilon^{ij}_{\ \  k} F_{ab}^{k}- 2(\gamma^{2}+1)\,K_{[a}^{i}K_{b]}^{ j}+  \nonumber\\
\!\!&-&\!\! \frac{2}{3} \frac{\alpha^2}{1+\gamma^2}\, \overline{K}_{(a}^{i} \overline{K}_{b)}^{ j} \Big) \,,
\eea
in which again we have assumed $\langle \mathcal{J}^i \rangle=\langle (\phi^{\dagger}\sigma^{i}\widetilde{D}_{a}\phi-
  \chi^{\dagger}\sigma^{i}\widetilde{D}_{a}\chi-c.c.)  \rangle=0$. The conditions imposed in order to recover the Ho\v{r}ava-Lifshitz scalar constraint are now $\lambda=3+ 2 \gamma^{2}$ and $3(\gamma^2+1)^2\!=\!\alpha^2$. Therefore the Immirzi parameter and $\lambda$ are now parametrized by the non-minimal coupling parameter $\alpha$ according to
\be
\gamma^2=\pm\frac{\alpha}{\sqrt{3}}-1\,, \quad {\rm and} \quad \lambda=1\pm \frac{2 \alpha}{\sqrt{3}}\,.
\ee
As a consequence, the condition to obtain the degenerate value $\lambda=1/3$, in order to derive exactly the quadratic Ho\v{r}ava-Lifshitz action in the Euclidean space
\bea \label{ok!}
 S\!\!=\!\!2 i \!\!\int \!\! dt d^3x \sqrt{q} N \Big\{ \frac{1}{{\kappa'}^2} K_{ij} G^{ijkl} K_{ij}  + \frac{{\kappa'}^2}{2 w^2} \widetilde{C}_{ij} \widetilde{C}^{ij}  \Big\}\,, 
\eea
can now be imposed, leading to
\be
\alpha=\mp\frac{1}{\sqrt{3}}\,.
\ee
This result sheds new light on the physical meaning of the dimensionless conformal coupling parameter $\lambda$, showing its connection with the non-minimal coupling parameter $\alpha$ that appears in (\ref{new action}). It is also interesting to note that any value $\lambda<3$ implies that only imaginary values are recovered for the Immirzi parameter. 


\section{Conclusions} 

\noindent In this paper we have shown how HL theory may be seen, in some situations, as the action of a fermionic aether in Ashtekar-like gravity in the presence 
of chiral spinor couplings. The torsion induced by the spinor generates an
extra term identical to that used in HL theory to break refoliation invariance.
This realization of Ho\v{r}ava gravity in the Ashtekar variables clarifies some open questions that were present
in the metric-variable formulation.  All of these issues are naturally connected by the condition of having a York-time, namely that the trace of the extrinsic curvature vanishes.  Once this condition is imposed the finiteness of the graviton is understood, since the Cotton tensor, which was assumed in the original Ho\v{r}ava formulation, gets related to the traceless part of the extrinsic curvature.  Furthermore, from the vanishing of the trace of the extrinsic curvature, we get a physical interpretation for the York-time \cite{Y2,Y1,Y3} as the fermionic electric charge density.  This identification can help us understand the issue of the loss of refoliation invariance as the physical fermionic aether which is the York-time, an issue we intend to pursue in future work.  

Given our results we can speculate further on why anisotropic scaling seems to 
lead to a renormalizable theory. The Einstein-Cartan-Kibble formulation of 
gravity is a gravity theory with torsion, but it is in fact equivalent to the 
torsion-free Einstein-Hilbert formulation if a four-fermion
(axial-axial) interaction is added to the latter.
It is well-known that four-fermion interactions are non-renormalizable.
Could it be that the non-renormalizable divergences they generate cancel the 
divergences associated with the usual perturbative treatment of gravity?
The equivalence exhibited in this paper would seem to imply that 
this is indeed the case;
however, it is far from trivial to prove it explicitly. If this is true
we can speculate further, and note that 
such a cancellation of divergences has a distinct flavour of supersymmetry 
about it. Could it
be that the fermionic degrees of freedom we are postulating result
from an underlying (super)-symmetry principle, capable of replacing
diffeomorphism or refoliation invariance? An answer in the affirmative
would explain many mysteries pertaining to HL theory, and why it works
so well. This intriguing possibility, however, 
remains a conjecture.

Finally, we should emphasize that in our formulation time diffeomorphism
invariance (refoliation invariance) is not explicitly broken. It is 
only spontaneously broken, as much 
as our Universe and the undeniable existence of a cosmological frame
are bound to
minimally break it.  This will necessarily soften the more unwanted 
implications of HL theory.
We conjecture, in particular, that a closer analysis of our model should
reveal an absence of the scalar graviton mode plaguing the theory. In addition
this seems to be possible without the need to introduce 
extra symmetries, such as in~\cite{HII} .
We defer to a future paper an extensive analysis of
this issue.

To summarize, Ho\v{r}ava's theory can be seen as a specific case 
of the covariant first-order gravity theory (Einstein-Cartan-Kibble-Holst).   
When the covariant theory is rewritten in Ashtekar variables, the 
imposition of the York-time yields the Ho\v{r}ava theory with the 
Cotton-tensor, in the presence of 
a fermion aether which breaks time-refoliation invariance.

\section{Acknowledgments} 
\noindent We would like to thank Steven Carlip, Pedro Ferreira, Tom Kibble,
Andrew Waldron and Tom Zlosnik for discussions and comments.

\end{document}